\documentclass[aps,prl,twocolumn,showpacs,superscriptaddress,groupedaddress]{revtex4}
\pdfoutput=1
\usepackage{hyperref}
\usepackage{epsfig}
\usepackage{amsmath}
\usepackage{graphicx}
\usepackage{wrapfig}
\usepackage{dcolumn}
\usepackage{bm}
\usepackage{amssymb}
\usepackage{titlesec}
\usepackage{mathtools}
\usepackage{relsize}
\usepackage{cleveref}
\usepackage[usenames,dvipsnames]{color}
\usepackage{ulem}

\begin{document}
%\title{\textcolor{blue}{Predicting} epidemic extinctions in heterogeneous networks}
\title{Epidemic extinction and control in heterogeneous networks}
\author{Jason Hindes}
\author{Ira B. Schwartz}
\affiliation{U.S. Naval Research Laboratory, Code 6792, Plasma Physics Division, Nonlinear Dynamical Systems Section, Washington, DC 20375}
\begin{abstract}

We consider epidemic extinction in finite networks with broad variation in local connectivity. Generalizing the theory of large fluctuations to random networks with a given degree distribution, we are able to predict the most probable, or optimal, paths to extinction in various configurations, including truncated power-laws. We find that paths for heterogeneous networks follow a limiting form in which infection first decreases in low-degree nodes, which triggers a rapid extinction in high-degree nodes, and finishes with a residual low-degree extinction. The usefulness of our approach is further demonstrated through optimal control strategies that leverage the dependence of finite-size fluctuations on network topology. Interestingly, we find that the optimal control is a mix of treating both high and low-degree nodes based on theoretical predictions, in contrast to methods that ignore dynamical fluctuations. %large-fluctuation theoretical predictions.  

%Interestingly, we find that neither targeting highly connected nodes for treatment, nor minimizing the endemic state, tend to minimize extinction times. 

%The usefulness of the approach is further demonstrated by developing optimal control strategies that leverage finite-size fluctuations. Interestingly, we find that neither targeting highly connected nodes for treatment, nor minimizing the endemic state, tend to minimize extinction times.      
\end{abstract}
\maketitle

Extinction of epidemics in finite networks is an important topic in population dynamics \cite{Anderson, Banavar}. Though many factors may contribute, such as environmental changes and social behavior, it has been demonstrated, and rigorously proven for finite populations, that internal fluctuations in a system's dynamics can organize in such a way to induce a large fluctuation along a most probable, or optimal, path to extinction \cite{Schwartz, Dykman, Meerson}. Such fluctuations to infection-free states have been studied extensively in well-mixed systems, including the role of vaccination and treatment programs in reducing the average time to extinction \cite{Schwartz, Billings}. Similarly, the most probable extinction paths have been found in networks with homogeneous degree, but the behavior appears to be independent of network topology, as in the well-mixed limit \cite{Lindley}.   

Somewhat separately, much work has been done in characterizing the deterministic dynamics, epidemic threshold, outbreak size distributions, small fluctuations, localization, and phases of epidemics in complex networks \cite{Pastor, Vespignani, Dorogovtsev, Newman, Munoz, Goltsev, Colizza}. Only very recently has there been progress in understanding the interplay between stochastic noise and network dynamics that can lead to large fluctuations and switching between states \cite{Motter}. However, very little is known about how internal noise inherent to epidemic extinction pertains in heterogeneous networks having vastly differing topologies. 

In this letter we construct and analyze the most probable path through heterogeneous networks to extinction. Novel in this work, is that we show the path has two primary forms, close-to and far-from the epidemic threshold. In the latter, we demonstrate an interesting multi-step structure in which low-degree node infections decrease first, followed by a quick, nearly complete extinction in high-degree nodes, and finishing with a low-degree extinction. The approach is then used to design a novel targeted optimal treatment strategy that can exponentially reduce extinction times, but does not trivially treat the most well connected nodes, minimize the epidemic size, nor maximize the number of treatments. Instead, the optimal control minimizes the ``action" associated with a transition to extinction with respect to the network topology. Such controls that manipulate finite-size fluctuations inherent in contact processes based on topology are also novel in the study of complex networks \cite{Billings, Motter}.

To understand how extinctions depend on topological heterogeneity, we consider the stochastic SIS model on uncorrelated random networks with a given degree distribution, $g_{k}$, where the degree, $k$, is the number of links of a node. Simple graphs with $N$ nodes can be generated from $g_{k}$ in several ways, for example with a configuration model, $CMN$\cite{Newman2}. Such networks are usefully represented by an adjacency matrix, $A$, where $A_{ij}$ is $1$ if nodes $i$ and $j$ are linked, and $0$ otherwise.  
%\Cor{Quenched networks} can be constructed with the configuration model ($CMN$), by first generating nodes, each with a number of link ``stubs" drawn from $g_{k}$, and then connecting pairs of ``stubs" to form links, chosen uniformly at random \cite{Newman2}. The adjacency matrix, $A$, for such a network, is defined such that, $A_{ij}$, is $1$ if nodes $i$ and $j$ are linked, and $0$ otherwise. In the latter, the network is assumed to fluctuate rapidly with respect to the SIS dynamics such that $A$ is given by the weighted average of $CMN$s \cite{Pastor, Vespignani, Colizza}.
In this representation a network's $SIS$ dynamics is captured by the states and transitions of its nodes, e.g., node $i$ is either infected, denoted $\nu_{i}\!=\!1$, or susceptible, $\nu_{i}\!=\!0$, and changes its state $\nu_{i}\!:\!0\rightarrow\!1$ 
with probability per unit time $\beta(1-\nu_{i})\sum_{j} A_{ij}\nu_{j}$, and $\nu_{i}\!:1\rightarrow0$ with probability per unit time $\alpha\nu_{i}$, where $\beta$ and $\alpha$ are known as the infection and recovery rates, respectively. 

In order to describe the dynamics given these reactions, it is useful to approximate $A_{ij}$ with its expectation value in an ensemble of networks, $\left<A_{ij}\right>$, which for uncorrelated networks takes the form, $\left<A_{ij}\right>= k_{i}k_{j}/(N\left<k\right>)$, in the limit of large $N$. 
%In order to describe the dynamics given these reactions, it is useful to approximate $A$ with the annealed limit, in which $A_{i j}$ approaches the probability that nodes $i$ and $j$ are neighbors, and is proportional to the product of their degrees, $k_{i}k_{j}$, or $A_{ij}\approx k_{i}k_{j}/(N\left<k\right>)$. 
This is known as the ``annealed" network approximation and represents a mean-field for heterogeneous networks \cite{Pastor,Vespignani, Dorogovtsev, Colizza}, though other techniques give similar results to those shown here \cite{Spectral}. Given this form, the state of the network can be described by the number of infected nodes with degree $k$, $I_{k}$, which has corresponding reactions and rates: $I_{k}\rightarrow I_{k}+1$ with rate of infection $w^{inf}_{k}(\bold{I})=\beta k(N_{k}-I_{k})\sum_{k'}k'I_{k'}/(N\left<k\right>)$, and $I_{k}\rightarrow I_{k}-1$ with recovery rate $w^{rec}_{k}(\bold{I})=\alpha I_{k}$, where $\bold{I}_{k}=\left<I_{k1},I_{k2},...,I_{k_{max}}\right>$, and $N_{k}=g_{k}N$.

Since the $SIS$ model is stochastic, associated with all network states is a probability distribution, $\rho(\bold{I},t)$, which for networks having a general degree distribution satisfies an approximate master equation:  
%Thus, the probability distribution, $\rho(\bold{I},t)$, for the $SIS$ model on networks having a general degree distribution satisfies an approximate master equation:  
\begin{align}
&\frac{\partial\rho}{\partial t}(\bold{I},t)=\sum_{k}w^{inf}_{k}(\bold{I}-\bold{1}_k)\rho(\bold{I}-\bold{1}_k,t)-w^{inf}_{k}(\bold{I})\rho(\bold{I},t) \nonumber \\
%&+\sum_{k}w^{rec}_{k}(\bold{I}+\bold{1}_k)\rho(\bold{I}+\bold{1}_k,t)-w^{rec}_{k}(\bold{I})\rho(\bold{I},t),
&+w^{rec}_{k}(\bold{I}+\bold{1}_k)\rho(\bold{I}+\bold{1}_k,t)-w^{rec}_{k}(\bold{I})\rho(\bold{I},t),
\label{eq:MasterEquation}
\end{align}
where $\bold{1}_{k}=\left<0{}\;_{k1},0{}\;_{k2},..,1{}\;_{k},..,0{}\;_{k_{max}}\right>$. We are interested in the behavior of Eq.(\ref{eq:MasterEquation}) for large, but finite networks. As customary, we assume $N$ is large and take the leading order in a $1/N$ expansion \cite{Schwartz, Dykman2}. This is similar to the Wentzel-Kramers-Brillouin, $WKB$, ansatz of quantum mechanics, where $1/N$ plays the role of Planck's constant in Shr{\"o}dinger's equation. In accordance with $WKB$, by writing $\rho(\bold{I},t)=e^{-NS(\bold{x},t)}$, where $\bold{x}=\bold{I}/N$, and taking the leading order in $1/N$, or $w_{k}(\bold{I}\pm\bold{1}_k)\approx w_{k}(\bold{I})$ and $\rho(\bold{I}\pm\bold{1}_{k},t)\approx e^{-NS(\bold{x})}e^{\mp\partial S/\partial x_{k}}$, we find a Hamilton-Jacobi equation, $\frac{\partial S}{\partial t}+H(\bold{x},\frac{\partial S}{\partial \bold{x}})=0$, where $S$ and $H$ are called the action and Hamiltonian,   
%It is known that for sufficiently large $N$, Eq.(\ref{eq:MasterEquation}) possesses quasi-stationary solutions about an endemic state with long-lived infection that decays into an extinct state over exponential time-scales \cite{Schwartz, Dykman2, Friedlin}. We therefore approximate such solutions with a $WKB$ ansatz; i.e., assume $\rho(\bold{I},t)=e^{-NS(\bold{x},t)}$, 
%where $\bold{x}=\bold{I}/N$, and take $w_{k}(\bold{I}\pm\bold{1}_k)\approx w_{k}(\bold{I})$ and $\rho(\bold{I}\pm\bold{1}_{k},t)\approx e^{-NS(\bold{x})}e^{\mp\partial S/\partial x_{k}}$. This gives 
%a Hamilton-Jacobi equation, $\frac{\partial S}{\partial t}+H(\bold{x},\frac{\partial S}{\partial \bold{x}})=0$, where $S$ and $H$ are called the action and Hamiltonian, 
respectively. As in classical mechanics, the Hamiltonian is a function of the coordinate, $\bold{x}$, and its conjugate momentum, $\bold{p}=\partial S/\partial \bold{x}$: 
\begin{align}
\!\!H(\bold{x},\bold{p})\!=\!\sum_{k}\!\!\Bigg[\!\beta k \big(g_{k}\!-\!x_{k}\big)\!\big(\!e^{p_{k}}\!-\!1\!\big)\!\!\sum_{k'}\!\frac{k'x_{k'}}{\left<k\right>}\!+\alpha x_{k}\big(\!e^{-p_{k}}\!-\!1\!\big)\!\Bigg]\!. 
\label{eq:Hamiltonian}
\end{align}
In this context, momenta behave as fluctuations on $\bold{x}$ -- describing both size and direction. 

It is convenient to analyze Eq.(\ref{eq:Hamiltonian}) by solving the canonical equations of motion: $\dot{x}_{k}=\partial H/\partial p_{k}$, $\dot{p}_{k}=-\partial H/\partial x_{k}$, in terms of the fraction of each degree class infected, $y_{k}\!=\!x_{k}/\!g_{k}$, the ratio $\beta/\alpha\!=\!\tilde{\beta}$, and the re-scaled time, $\tau\!=\!\alpha t$:
\begin{align}
\label{eq:EOM}
\dot{y}_{k}&=\!\tilde{\beta}k(1-y_{k})e^{p_{k}}\!\sum_{k'}\!\frac{k'g_{k'}}{\left<k\right>}y_{k'}-y_{k}e^{-p_{k}},\\
\dot{p}_{k}&= \!\tilde{\beta}k\!\sum_{k'}\!\frac{k'g_{k'}}{\left<k\right>}\!\Bigg[\!y_{k'}\big(\!e^{p_{k}}\!-\!1\!\big)\!-\!\big(\!1\!-\!y_{k'}\!\big)\!\big(\!e^{p_{k'}}\!-\!1\!\big)\!\Bigg]\!\!-\!e^{\!-p_{k}}\!\!+\!1.\!\nonumber
\end{align}

\begin{figure}[h]
\includegraphics[scale=0.23]{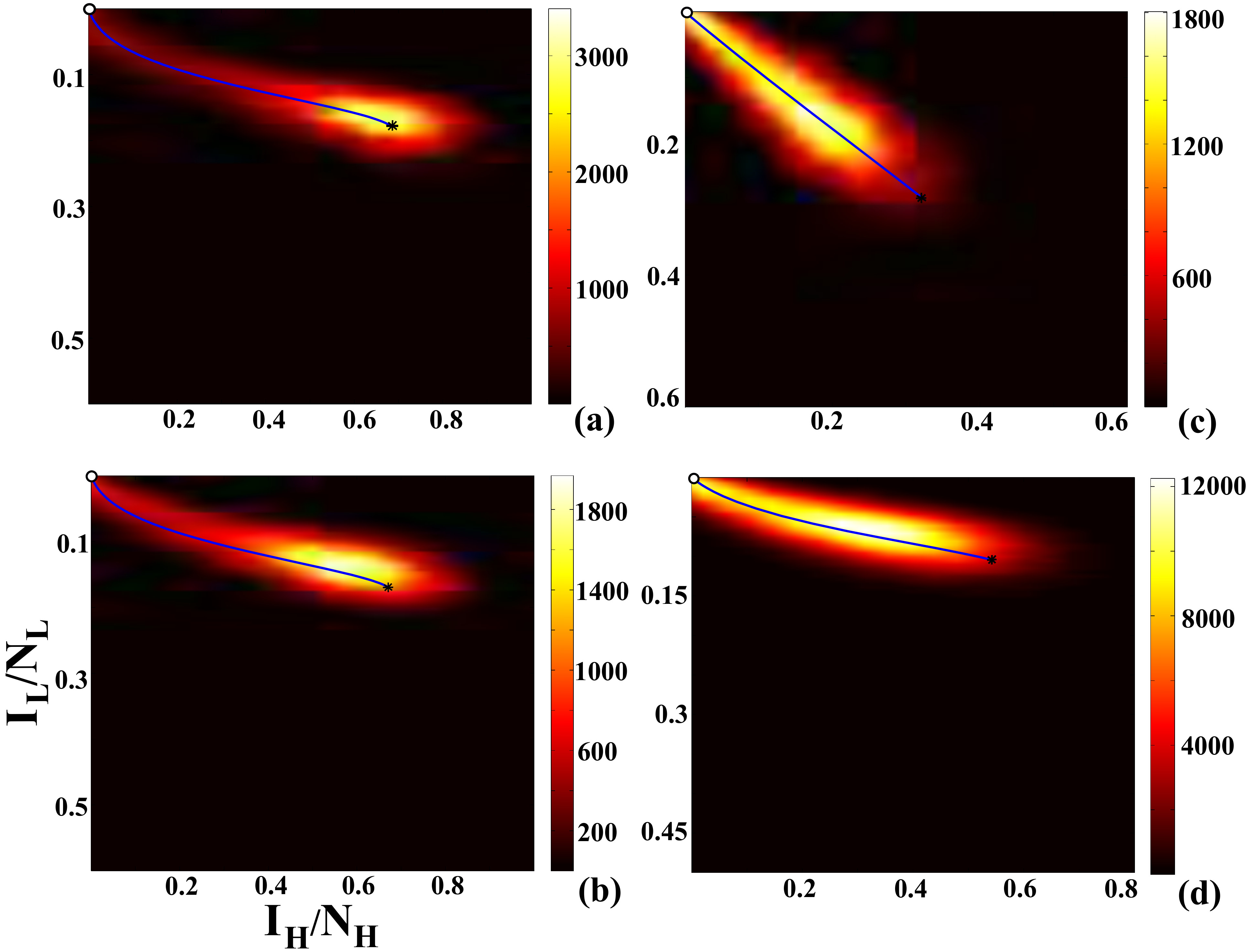}
\caption{Density (unnormalized) of $1000$ simulations projected into the fraction of infected high (H) and low-degree (L) nodes. Predicted paths are shown in blue from the endemic state($*$) to extinction($\circ$). (a) a network with $A_{ij}=k_{i}k_{j}/(N\left<k\right>)$, $N\!=\!300$, and $\tilde{\beta}\!=\!0.096$, and with two degree classes, $k_{i}\in\{5,50\}\!$; $k\!=\!50$-nodes occupying $\!10\%\!$ of the network. (b) a corresponding $CMN$. (c) a $CMN$ with $N=350$, $\tilde{\beta}=0.092$, and $g_{k}\!=\!e^{-16}16^{k}/k!$ where high-degree nodes have $16\!\leq\!k\!\leq\!18$, and low-degree have $13\!\leq\!k\!\leq\!15$. (d) a $CMN$ with $N\!=\!600$, $\tilde{\beta}\!=\!0.038$, and $g_{k}\!=\!k^{-2.5}\!/\!\sum_{k'=10}^{300}k'^{-2.5}$ where high-degree nodes have $70\!\leq\!k\!\leq\!235$, and low-degree have $10\!\leq\!k\!\leq\!12$.}
\label{fig:Gill}
\end{figure}

Of interest are particular solutions of Eq.(\ref{eq:EOM}) which correspond to network trajectories that remain near an endemic state for some time, and then decay into an extinct state, with no more infectious nodes. When the two states are well separated, the distribution $\rho(\bold{I},t)$ is quasi-stationary, or in the $WKB$ ansatz, $\frac{\partial S}{\partial t}=H=0$:
\begin{align}
S=\int{\big[\bold{p}\cdot\dot{\bold{x}}-H\big]dt}=\sum_{k}g_{k}\int{p_{k}dy_{k}}. 
\label{eq:Action}
\end{align}
This suggests that we look for solutions of Eq.(\ref{eq:EOM}) in the form of heteroclinic paths connecting two saddle-point equilibria: from an endemic fixed-point, $y_{k}^{*}\!\!=\!\!1/(1\!+\!1/(Y\tilde{\beta} k)), \; p_{k}\!=\!0$, to extinction with non-zero momentum, $y_{k}\!=\!0, \; p_{k}^{*}\!=\!-\!\ln(1\!+\!\tilde{\beta}k(1\!-\!P))$ \cite{Meerson, Schwartz, Elgart}. The functions $Y$ and $P$ depend on $\tilde{\beta}$, with $(Y,P)\!\rightarrow\!(0,1)$ as $\tilde{\beta}\left<k^{2}\right>\!\big/\!\left<k\right>\!\equiv\!R_{0}\!\rightarrow\!1$, and $(Y,P)\!\rightarrow\!(1,0)$ as $\tilde{\beta}\!\rightarrow\!\infty$. Importantly, because such paths extremize their action, they extremize their probability, and therefore correspond to most probable paths through a network to extinction \cite{Dykman2, Friedlin}.
%For the quasi-stationary distribution, $\frac{\partial S}{\partial t}=H=0$, the optimal path from the endemic state to extinction is one that extremizes the action\cite{Friedlin}: 
%\begin{align}
%S=\int{\big[\bold{p}\cdot\dot{\bold{x}}-H\big]dt}=\sum_{k}g_{k}\int{p_{k}dy_{k}},
%\label{eq:Action}
%\end{align}
%with the average time to extinction scaling as $\left<T\right>\!\sim\!e^{NS}$. The path is a heteroclinic orbit connecting two equilibria of Eq.(\ref{eq:EOM}): from an endemic fixed-point, $y_{k}^{*}\!\!=\!\!1/(1\!+\!1/(Y\tilde{\beta} k)), \; p_{k}\!=\!0$, to extinction with non-zero momentum, $y_{k}\!=\!0, \; p_{k}^{*}\!=\!-\!\ln(1\!+\!\tilde{\beta}k(1\!-\!P))$ \cite{Meerson, Schwartz, Elgart}. The functions $Y$ and $P$ depend on $\tilde{\beta}$, with $(Y,P)\!\rightarrow\!(0,1)$ as $\tilde{\beta}\left<k^{2}\right>\!\big/\!\left<k\right>\!\equiv\!R_{0}\!\rightarrow\!1$, and $(Y,P)\!\rightarrow\!(1,0)$ as $\tilde{\beta}\!\rightarrow\!\infty$. 
Fig.(\ref{fig:Gill}) shows comparisons between projections of pre-history trajectories to extinction from stochastic simulations and optimal paths of Eq.(\ref{eq:EOM}) for several network configurations computed with the iterative action minimizing method (IAMM) \cite{Lindley2}.

In general, the average time to reach extinction, $\left<T\right>$, will depend on $\tilde{\beta}$ and network properties in complicated ways \cite{Doering,Holme}. However, for sufficiently large $N$, the transition is an exponential process with a rate proportional to the probability, and therefore $\left<T\right>\!\sim\!e^{NS}$. The network action, Eq.(\ref{eq:Action}), is thus central to understanding the dependencies of extinction times on network topology.

\begin{figure}[h]
\raggedright{\includegraphics[scale=0.242]{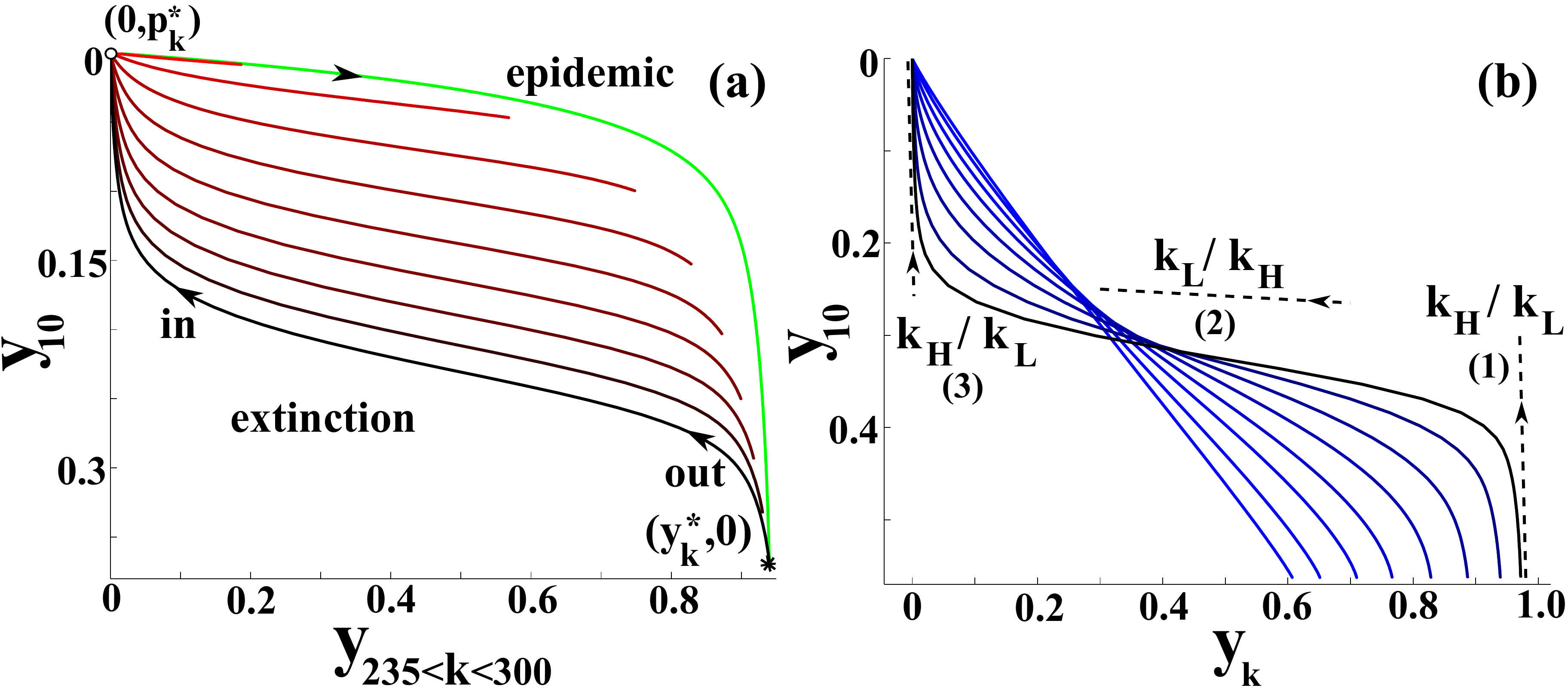}}
\caption{(a) Projections of the optimal paths for truncated power-law (see Fig.\ref{fig:Gill}(d)) shown for increasing $R_{0}\!\in\![1.1,5.1]$ (red$\rightarrow$black, weak$\rightarrow$strong) in steps of $0.5$, compared with the path into the endemic state for $R_{0}\!\!=\!\!5.1$ (green). Arrows indicate direction in time. (b) Projections into $y_{k}$ for the same distribution with $R_{0}\!=\!9$ and $17$ bins\cite{Binning}, shown for bins with increasing $k$: $\{k\!=\!12, 14\!\leq \!k\! \leq\!15, 18\!\leq \!k\! \leq\!20, 24\!\leq \!k\! \leq\!27, 34\!\leq \!k\! \leq\!41, 53\!\leq \!k\! \leq\!69, 97\!\leq \!k\!\leq\!143, 236\!\leq \!k\! \leq\!300\}$ (blue$\rightarrow$black), and compared with the predicted scaling for the highest bin (dashed lines).}
\label{fig:Paths}
\end{figure}

Qualitatively, we find two important parameter regions to consider. First, close to the epidemic threshold when $R_{0}\!-\!1\!\!\gtrsim\!0$ (which we call ``weak"), the paths to extinction are approximately linear from $(y_{k}^*,0)$ to $(0,p_{k}^{*})$. The explicit form can be seen by expanding the equilibria in powers of $R_{0}\!-\!1$, which gives to first order, $y_{k}^{*}\approx k\!\left<k^{2}\right>\!(R_{0}-1)/\!\left<k^{3}\right>=-p_{k}^{*}$, implying that the endemic state and momentum at extinction are simply proportional to degree when the infection is weak. This is intuitive since in the weak limit high-degree nodes drive the epidemic, when only their local reproductive numbers are sufficient to spread infection, ($\tilde{\beta}k\!>\!1$), and therefore must recover disproportionately without reinfection in order for extinction to occur. Paths near the weak limit can be seen in Fig.\ref{fig:Gill}(b) and (d) where $R_{0}\!=\!1.6$ and $R_{0}\!=\!2.0$ respectively, and in Fig.\ref{fig:Paths}(a)-(red). 

Also for weak infection, the action along the path from Eq.(\ref{eq:Action}) is therefore:
\begin{align}
S_{weak}=\frac{\;\;\left<k^2\right>^3}{2\left<k^{3}\right>^{2}}\big(R_{0}-1\big)^{2}+\mathcal{O}\big(R_{0}-1\big)^{3},
\label{eq:Weak}
\end{align}
which depends on the distance from the epidemic threshold and a non-trivial topological factor that generally decreases with increased broadness in the degree distribution. In contrast, in the well-mixed limit (corresponding to the simple complete graph) the action only depends on $R_{0}$ \cite{Schwartz, Doering}; The predicted reduction in extinction times with topological fluctuations is intuitive, since for very heterogeneous networks only a small fraction of highly connected nodes must recover without reinfection, compared with most nodes in networks where nodes are topologically similar. 

On the other hand if most nodes can propagate infection, $\tilde{\beta}\!\left<k\right>\!\!\gg\!1$ (which we call ``strong"), then the interplay between degree classes and the path to extinction are more complicated as the global dynamical structure of the path becomes apparent. However, we find that a limiting form emerges when comparing the dynamics of low and high-degree nodes by which the path can be described in multiple steps.

Since in the strong limit most nodes will be infected in the endemic state, it is very improbable that high-degree nodes can recover without being reinfected, and thus we expect infection must first disproportionately decrease in low-degree nodes. We can extract the form of this step, by analyzing the unstable eigen-mode of the endemic equilibrium: $(y_{k},p_{k})=(y_{k}^{*}+\epsilon_{k}^{(1)},\mu_{k}^{(1)})$, to linear order and studying the asymptotic scaling of $(\epsilon_{k}^{(1)},\mu_{k}^{(1)})$ for large $\tilde{\beta}k$. Inserting these assumptions into Eq.(\ref{eq:EOM}), we find that $(\epsilon_{k}^{(1)},\mu_{k}^{(1)})$ must satisfy an eigenvalue equation for the rate $\lambda^{(1)}$:
%\vspace{-1.0cm}
\begin{equation*}
%\vspace{-0.8cm}
\nonumber
%\Bigg[\!\lambda^{(1)}+\!1\!+\frac{\tilde{\beta}^{2}\!\left<k\right>\!k\!}{\tilde{\beta}\!\left<k\right>\!-1\!}\Bigg]\!\epsilon_{k}^{(1)}\!-2\Bigg[\frac{\tilde{\beta}\!\left<k\right>}{\tilde{\beta}\!\left<k\right>\!-1\!}\Bigg]\!\mu_{k}^{(1)}\approx\!\sum_{k'}\!\frac{k'g_{k'}}{\left<k\right>}\epsilon_{k'}^{(1)}. \nonumber
%\!\Bigg[\!\lambda^{(1)}+1+\frac{k\big(\tilde{\beta}\!\left<k\right>\!-\!1\!\big)}{\left<k\right>}\Bigg]\!\epsilon_{k}^{(1)}\!-2\Bigg[\frac{\tilde{\beta}\!\left<k\right>\!-1\!}{\tilde{\beta}\!\left<k\right>}\Bigg]\!\mu_{k}^{(1)}\approx\!\sum_{k'}\!\frac{k'g_{k'}}{\left<k\right>}\epsilon_{k'}^{(1)}. \nonumber\label{eq:StepOne} 
\!\!\Bigg[\!\lambda^{(1)}+1+\frac{k\big(\tilde{\beta}\!\left<k\right>\!-\!1\!\big)}{\left<k\right>}\!\Bigg]\!\epsilon_{k}^{(1)}\!-\Bigg[\!2-\frac{1}{\tilde{\beta}k}-\frac{1}{\tilde{\beta}\!\left<k\right>}\!\Bigg]\!\mu_{k}^{(1)}\!\!\approx\!\!\sum_{k'}\!\frac{k'g_{k'}}{\left<k\right>}\epsilon_{k'}^{(1)}\!\!. \nonumber\label{eq:StepOne} 
\end{equation*}

%\vspace{-1.2cm}
\noindent Since $\lambda^{(1)}$ and the sum are $k$-independent, it must be that for large $\tilde{\beta}k$, we have $\epsilon_{k}^{(1)}/\epsilon_{k'}^{(1)}\sim k'/k$ (the relative decrease in infection shown in Fig.\ref{fig:Paths}(b)-(1)), with relative momenta initially tending to a constant. 

In the second step (Fig.\ref{fig:Paths}(b)-(2)), the small build-up of momenta for high-degree nodes becomes rapid as the $k$-dependent contribution to $\dot{p}_{k}$ approaches a maximum along the path. In analogy with mechanics, this can be thought of as the network's contribution to the ``force" on $y_{k}$,
%\begin{equation*}
%F_{k}\!=\!\sum_{k'}\!\frac{k'g_{k'}}{\left<k\right>}\!\Bigg[\!y_{k'}\big(\!e^{p_{k}}-1\!\big)\!-\!\big(\!1-y_{k'}\!\big)\!\big(\!e^{p_{k'}}-1\!\big)\!\Bigg]\!\!\rightarrow\! \text{min}[F_{k}], \nonumber
%\label{eq:StepOne} 
%\end{equation*}
which near its maximum quickly ``pushes" $-p_{k}$ from near zero to its maximum, $-p_{k}^{*}$. On the other hand, since $y_{k}$ and $p_{k}$ decreases along the path, by inspecting $\dot{y}_{k}$, we can find an upper bound for $-\dot{y}_{k}$, $\text{max}[-\dot{y}_{k}]< \text{max}[-y_{k}e^{-p_{k}}]<-y_{k}^{*}e^{-p_{k}^{*}}$, because the $k$-dependent contribution to $\dot{y}_{k}$ is positive. Since $y_{k} \approx y_{k}^{*}$ until $-p_{k}$ differs from zero, as the force approaches its maximum, $-\dot{y}_{k}$ can be approximated by the upper bound, giving the scaling for large $\tilde{\beta}k$: $\epsilon_{k}^{(2)}/\epsilon_{k'}^{(2)}\sim k/k'$.

%Also, we can see from Eq.(\ref{eq:EOM}) that only when $-p_{k}$ is large does $y_{k}$ decrease significantly, implying that $y_{k}$ remains near the endemic state until the large momentum of this step is reached. This suggests that we can approximate the average speed of step-two with the upper bound of the maximum possible speed, $\text{max}[-\dot{y}_{k}]\lesssim y_{k}^{*}e^{-p_{k}^{*}}$ giving the scaling for large $\tilde{\beta}k$: $\epsilon_{k}^{(2)}/\epsilon_{k'}^{(2)}\sim k/k'$. %, \textcolor{blue}{(what is the force? balance of terms? smooth paragraph)}. 

In the last step (Fig.\ref{fig:Paths}(b)-(3)), we expect to have a final decrease in low-degree node infections in a background of very small numbers of infected high-degree nodes, since the latter were rapidly depleted in the second step. The scaling can be found by analyzing the extinct state's stable eigen-mode: $(y_{k},p_{k})=(\epsilon_{k}^{(3)},p_{k}^{*}+\mu_{k}^{(3)})$, which gives an eigenvalue equation for the rate $\lambda^{(3)}$, 
\begin{equation*}
\!\Bigg[\!1+\frac{1}{\tilde{\beta}k}-\frac{1}{\tilde{\beta}\!\left<k\right>}\Bigg]\!\Bigg[\!\lambda^{(3)}+1+\frac{k\big(\tilde{\beta}\!\left<k\right>\!-\!1\!\big)}{\left<k\right>}\!\Bigg]\epsilon_{k}^{(3)}\!\approx\!\sum_{k'}\!\frac{k'g_{k'}}{\left<k\right>}\epsilon_{k'}^{(3)}\!, \nonumber
%\Big[\lambda^{(3)}+\tilde{\beta}k+1\Big]\epsilon_{k}^{(3)}\approx\!\Bigg[\frac{\tilde{\beta}k}{\tilde{\beta}k+1}\Bigg]\!\sum_{k'}\!\frac{k'g_{k'}}{\left<k\right>}\epsilon_{k'}^{(3)}, \nonumber
\label{eq:StepThree} 
\end{equation*} 
that in the limit of large $\tilde{\beta}k$ implies $\epsilon_{k}^{(3)}/\epsilon_{k'}^{(3)}\sim k'/k$. Examples are shown in Fig.\ref{fig:Paths}(a)(black) and Fig.\ref{fig:Paths}(b)(black); in the former, the strong limit scaling starts to be visible for $R_{0}\!\gtrsim3$. Also, Fig.\ref{fig:Paths}(a) shows the significant qualitative difference between the epidemic path toward the endemic state (green), which has been well studied, and the optimal path to extinction \cite{Pastor}.
%\begin{figure}[h]
%\includegraphics[scale=0.285]{Paths2.pdf}
%\caption{(a) Projections of the optimal paths for truncated power-law (see Fig.\ref{fig:Gill}(d)) shown for increasing $R_{0}\!\in\![1.1,5.1]$ (red$\rightarrow$black, weak$\rightarrow$strong) in steps of $0.5$, compared with the path into the endemic state for $R_{0}\!\!=\!\!5.1$ (green). Arrows indicate direction in time. (b) Projections into $y_{k}$ for the same distribution with $R_{0}\!=\!9$ and $17$ bins\cite{Binning}, shown for bins with increasing $k$: $\{k\!=\!12, 14\!\leq \!k\! \leq\!15, 18\!\leq \!k\! \leq\!20, 24\!\leq \!k\! \leq\!27, 34\!\leq \!k\! \leq\!41, 53\!\leq \!k\! \leq\!69, 97\!\leq \!k\!\leq\!143, 236\!\leq \!k\! \leq\!300\}$ (blue$\rightarrow$black), and compared with the predicted scaling for the highest bin. (c) Action versus  the fraction of infected high-degree nodes treated in a bimodal network (see Fig.\ref{fig:Gill}(a)-(b)) and increasing treatment rate, $\gamma$ (red$\rightarrow$black): $\tilde{\beta}=0.275$. The inset shows the extinction times for a $CMN$ with $N=200$.}
%\label{fig:Paths}
%\end{figure} 

In addition to a theoretical interest in the geometry of the optimal path through a network, it is also practically interesting, because extinction times scale exponentially with the action \cite{Schwartz}. Since the action depends nontrivially on network topology, eg. Eq.(\ref{eq:Weak}), we suggest exploiting topology as a basis for optimal epidemic control strategies in finite networks, with the goal of minimizing extinction times \cite{Billings, Motter,Drakopoulos}. We illustrate the approach with a random treatment procedure for infected nodes with degree $k$, such that they recover with an increased rate, $\alpha+\gamma w_{k}$, where $\gamma$ is the overall treatment rate, and $w_{k}$ is a targeting fraction of the infected population with degree $k$: $\sum_{k}w_{k}=1$.
\begin{figure}[h]
\includegraphics[scale=0.275]{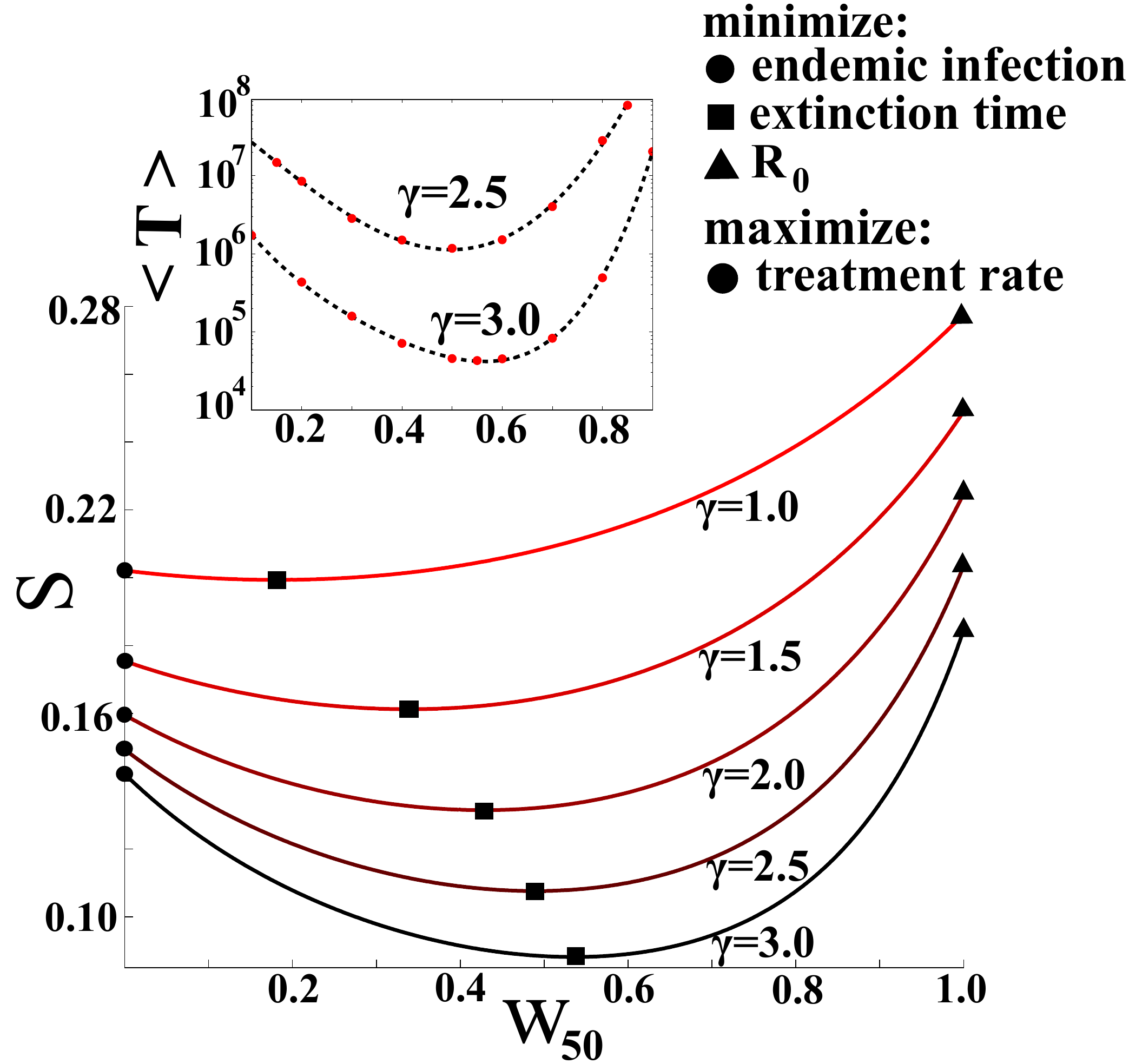}
\caption{Action versus  the fraction of infected high-degree nodes treated in a bimodal network (see Fig.\ref{fig:Gill}(a)-(b)) and increasing treatment rate, $\gamma$ (red$\rightarrow$black): $\tilde{\beta}\!=\!0.275$. The inset shows the extinction times for a $CMN$ with $N\!=\!200$.}
\label{fig:Control}
\end{figure} 

In the weak limit we expect optimal treatment to favor large $k$ (similar to targeted immunization), since $y_{k}^{*}$ and $p_{k}^{*}$ are proportional to $k$ \cite{Pastor2}. However in the strong limit, treating low-degree nodes close to $\left<k\right>$ will tend to decrease $S$, since their numbers must be lowered in the first step, before momenta differ significantly from zero. In intermediate cases, we expect a mixed strategy to minimize $S$. Treatment results are shown in Fig.\ref{fig:Control} for a simple bimodal network with two degree classes for clarity, as a function of the targeting fraction for high-degree nodes, $w_{50}$.     

Interestingly, we find that choosing the optimal $w_{k}$ for the bimodal network can result in a nearly $50\%$ decrease in the network action, implying an enormous reduction in extinction times, i.e., $\left<T\right>\!\!\rightarrow\!\!\left<T\right>^{1\!/2}$ (Fig.\ref{fig:Control}-inset). Furthermore, we note that in Fig.\ref{fig:Control} the size of the endemic state is minimized when $w_{50}\!=\!0$ (cirlces), the equilibrium treatment rate, $\gamma\sum_{k}w_{k}g_{k}y_{k}^{*}$, is maximized when $w_{50}\!=\!0$ (circles), and $R_{0}$ is minimized when $w_{50}\!=\!1$ (triangles), but none correspond to the minimum extinction time control (squares) \cite{R0}. The example demonstrates that designing optimal controls intended to drive epidemics to extinction in finite networks cannot be found from the intuitive results of the deterministic limit alone, $p_{k}\!=\!0$, but by targeting the network's components in such a way as to minimize the network's action.

%This demonstrates that control efforts that leverage stochasticity in finite networks,  
%1. control leverages a purely stochastic effect which does not agree with 
%
%This demonstrates that knowing the optimal paths for finite heterogeneous networks is very useful for designing controls intended   
%to drive epidemics to extinction,
 %unlike homogeneous systems where the lowest action would result from decreasing the overall epidemic size, finite heterogeneous networks have non-trivial optimal paths on which the action depends, and therefore knowledge of the path is necessary for designing efficient controls intended to drive epidemics to extinction. \\ 

In conclusion, we have considered how fluctuations in the SIS model produce extinctions from internal noise in finite heterogeneous networks, and found that the process is captured by a most probable path. We were able to construct paths by combining the theory of rare events and random networks with a general degree distribution, and  predict important consequences, such as the exponential decrease in extinction times with topological variation, as well as the multi-step scaling of extinction through nodes with very different degree. Furthermore, we demonstrated how the theory can be used to manipulate fluctuations for optimal network control, producing exponential decreases in extinction times with a simple treatment strategy that minimized the action by leveraging its dependence on topology. Our theoretically predicted results were confirmed by simulations over large parameter ranges and different network topologies. 

Lastly, we suggest the theoretical approach can be tailored to more arbitrary weighted networks and general epidemic processes that would allow one to predict the paths to extinction through real networks \cite{Anderson, Pastor, Goltsev,Spectral}. The specific formalism presented here could be augmented, in addition, to include degree correlations that may amplify or reverse the patterns described in interesting ways depending on the network assortativity.   

%\\and could be applied to more general epidemic processes \cite{Anderson, Pastor, Goltsev,Spectral}.}  

%for bacterial diseases, for example, where the $SIS$ model is appropriate \cite{Anderson, Goltsev}.   
%\section*{\label{sec:Ack}ACKNOWLEDGMENTS} 
We are grateful to Luis Mier-y-Teran Romero, D. J. Schneider, B. S. Lindley, C. R. Myers, and L. B. Shaw for useful discussions. J. H. is a National Research Council postdoctoral fellow. I.B.S was supported by the U.S. Naval Research Laboratory funding (N0001414WX00023) and Office of Naval Research (N0001414WX20610).


\begin{thebibliography}{9}
\bibitem{Anderson} R. M. Anderson and R. M. May, {\it Infectious Diseases of Humans} (Oxford University Press, 1991).
\bibitem{Banavar} J. R. Banavar and A. Maritan, Nature {\bf 460}, 334 (2009). 
\bibitem{Schwartz} M. I. Dykman, I. B. Schwartz, and A. S. Landsman, Phys. Rev. Lett. {\bf 101}, 078101 (2008).  
\bibitem{Dykman} M. Khasin and M. I. Dykman, Phys. Rev. Lett. {\bf 103}, 068101 (2009).
\bibitem{Meerson} A. Kamenev, B. Meerson, and B. Shklovskii, Phys. Rev. Lett. {\bf 101}, 268103 (2008). 
\bibitem{Billings} L. Billings, L. Mier-y Teran Romero, B. S. Lindley, and I. B. Schwartz, PLoS One {\bf 8}, e70211 (2013).
\bibitem{Lindley} B. S. Lindley, L. B. Shaw, and I. B. Schwartz, Europhys. Lett. {\bf 108}, 58008 (2014). 
\bibitem{Pastor} R. Pastor-Satorras, C. Castellano, P. Van Mieghem, and A. Vespignani, Rev. Mod. Phys. {\bf 87}, 925 (2015).
\bibitem{Vespignani} R. Pastor-Satorras and A. Vespignani, Phys. Rev. Lett. {\bf 86}, 3200 (2001).
\bibitem{Dorogovtsev} S. N. Dorogovtsev, A. V. Goltsev, and J. F. F. Mendes, Rev. Mod. Phys. {\bf 80}, 1275 (2008).
\bibitem{Newman} M. E. J. Newman, Phys. Rev. E {\bf 66}, 016128 (2002).
\bibitem{Munoz} M. A. Mu$\tilde{\text{n}}$oz, R. Juh\'{a}sz, C. Castellano, and G. \'{O}dor, Phys. Rev. Lett. {\bf 105}, 128701 (2010).
\bibitem{Goltsev} A. V. Goltsev, S. N. Dorogovtsev, J. G. Oliveira, and J. F. F. Mendes, Phys. Rev. Lett. {\bf 109}, 128702 (2012). 
\bibitem{Colizza} E. Valdano, L. Ferreri, C. Poletto, V. Colizza, Phys. Rev. X {\bf 5}, 021005 (2015).  
\bibitem{Motter} D. K. Wells, W. L. Kath, and A. E. Motter, Phys. Rev. X {\bf 5}, 031036 (2015).
\bibitem{Newman2} M. E. J. Newman, {\it{Networks: An Introduction}} (Oxford University Press, 2010).
\bibitem{Spectral} E.g., analogous results can be derived by directly approximating $A$, in which case degrees and their statistical moments are replaced by eigenvector centralities and eigenvalues of $A$ \cite{Pastor,Goltsev}.  
%Spectral approaches can provide quantitative improvements in accuracy \cite{Pastor,Goltsev}, e.g, in networks with large spectral gaps.
\bibitem{Dykman2} M. I. Dykman, E. Mori, J. Ross, and P. M. Hunt, J. Chem. Phys. {bf 100}, 5735 (1994).  
\bibitem{Elgart} V. Elgart and A. Kamenev, Phys. Rev. E {\bf 70}, 041106 (2004). 
\bibitem{Friedlin} M. I. Friedlin and A.D. Wentzell, {\it Random Perturbations of Dynamical Systems} (Springer-Verlag, New York, 1998), 2nd ed.. 
\bibitem{Lindley2} B. S. Lindley and I. B. Schwartz, Physica D {\bf255}, 22 (2013).
\bibitem{Doering}C.R. Doering, K. V. Sargsyan, and L. M. Sander, Multiscale Model. Simul. {\bf3}(2), 283 (2005).
\bibitem{Holme} P. Holme, PLoS One {\bf 8}(12), e84429 (2013).
\bibitem{Drakopoulos} K. Drakopoulos, A. Ozdaglar, and J. N. Tsitsiklis, IEEE Trans. Netw. Sci. Eng. {\bf 1}(2), 67 (2014). 
\bibitem{Pastor2} R. Pastor-Satorras and A. Vespignani, Phys. Rev. E {\bf 65}, 036104 (2002).
\bibitem{Binning} Since the path is hyperbolic and power-laws have many degree classes, it is useful to apply a binning procedure that reduces the dimension for the IAMM. Above, we have uniformly binned the distribution $kg_{k}/\!\left<k\right>$, and replaced $k$ with the average in each bin. This was chosen because it gives accurate approximations for $\left<k^{2}\right>\!\big/\!\left<k\right>$. Also, $k_{max}$ was chosen small enough such that at least $\mathcal{O}(1)$ nodes could be expected in its bin given 
 $N$.
 \bibitem{R0} With treatment, $R_{0}-1$ is the distance from the critical point and satisfies: $\!\mathlarger{\sum_{k}}\dfrac{kg_{k}}{\left<k\right>}\dfrac{\tilde{\beta}k}{R_{0}+\gamma w_{k}/\alpha}=1$.  
\end{thebibliography}
\end{document}